\documentclass[conference]{IEEEtran}

\IEEEoverridecommandlockouts
\usepackage{spconf,amsmath,graphicx,cite,color,balance,amsfonts,subfigure,float}
\usepackage{stfloats}
\usepackage{cite}
\usepackage[T1]{fontenc}

\usepackage{stmaryrd}
\usepackage{amssymb}
\usepackage{mathrsfs}
\usepackage{amsfonts}  
\usepackage{algorithm}
\usepackage{algorithmic}
\usepackage{cuted}
\usepackage{balance}
\usepackage{tabularx}
\usepackage{booktabs}

\usepackage{booktabs}
\usepackage{stfloats}
\usepackage{multirow}

\usepackage{dsfont}
\usepackage{esint}
\usepackage{url}
\usepackage{subfig}
\usepackage{float}

\usepackage{enumitem}
\usepackage{tikz}
\usepackage{caption} 
\usepackage{tabularx}
\usepackage{relsize}\usepackage{subdepth}\allowdisplaybreaks

\newcommand{\beq}{\begin{equation}}
\newcommand{\eeq}{\end{equation}}
\newcommand{\beqn}{\begin{eqnarray}}
\newcommand{\eeqn}{\end{eqnarray}}

\makeatletter 
\newcommand\semiLarge{\@setfontsize\semiLarge{10.72}{15.38}}
\makeatother

\usepackage{etoolbox,balance}

\patchcmd{\thebibliography}{\section*{\refname}}{}{}{}
\vspace{-4.5em}
\title{IHT-Inspired Neural Network for Single-Snapshot DOA Estimation with Sparse Linear Arrays}
\vspace{-1.2em}
\name{Yunqiao~Hu and Shunqiao~Sun
\vspace{-0.5em} 
\thanks{This work has been funded in part by U.S. National Science Foundation (NSF) under Grants CCF-2153386.}
}
\address{\hspace{-0.75em} \semiLarge
Department of Electrical and Computer Engineering, The University of Alabama, Tuscaloosa, AL, USA  
}

\begin{document}
\begin{sloppypar}
\ninept
\maketitle

\begin{abstract}
Single-snapshot direction-of-arrival (DOA) estimation using sparse linear arrays (SLAs) has gained significant attention in the field of automotive MIMO radars. This is  due to the dynamic nature of automotive settings, where multiple snapshots aren't accessible, and the importance of minimizing hardware costs. Low-rank Hankel matrix completion has been proposed to interpolate the missing elements in SLAs. However, the solvers of matrix completion, such as iterative hard thresholding (IHT),  heavily rely on expert knowledge of hyperparameter tuning and lack task-specificity. Besides, IHT involves truncated-singular value decomposition (t-SVD), which has high computational cost in each iteration. In this paper, we propose an IHT-inspired neural network for single-snapshot DOA estimation with SLAs, termed IHT-Net. We utilize a recurrent neural network structure to parameterize the IHT algorithm. Additionally, we integrate shallow-layer autoencoders to replace t-SVD, reducing computational overhead while generating a novel optimizer through supervised learning. IHT-Net maintains strong interpretability as its network layer operations align with the iterations of the IHT algorithm. The learned optimizer exhibits fast convergence and higher accuracy in the full array signal reconstruction followed by single-snapshot DOA estimation. Numerical results validate the effectiveness of the proposed method.
\end{abstract}
\begin{keywords}
Sparse linear array, matrix completion, iterative hard thresholding, deep neural networks, single snapshot, direction-of-arrival estimation
\end{keywords}

\section{Introduction}
Millimeter wave (mmWave) radar is highly reliable in various weather environments and antennas can be fit in a small form factor to provide high angular resolution, enhancing the environment perception capabilities. Compared with LiDAR, mmWave radar is a more cost-effective solution, making it crucial for autonomous driving\cite{SUN_SPM_Feature_Article_2020,sun20214d,Markel_book_2022}. Benefiting from multiple-input multiple-output (MIMO) radar technology, mmWave radars can synthesize virtual arrays with large aperture sizes using a small number of transmit and receive antennas\cite{SUN_SPM_Feature_Article_2020}. To further reduce the hardware cost, sparse arrays synthesized by MIMO radar technology has been widely adopted in automotive radar \cite{sun2020sparse, sun20214d, sun2023fast}.

Direction-of-arrival (DOA) estimation is one significant task for automotive radar. Classic subspace-based DOA estimation algorithms such as MUSIC \cite{schmidt1986multiple} and ESPRIT \cite{roy1989esprit} require multiple snapshots to yield accurate DOA estimates. However, in highly dynamic automotive scenarios, only limited radar snapshots or even just a single snapshot are available for DOA estimation. Consequently, research on single-snapshot DOA methods with sparse arrays is of significant importance.
The challenges associated with single-snapshot DOA with sparse arrays are the high sidelobes and the reduction of signal-to-noise ratio (SNR), both of which may cause errors and ambiguity in estimation\cite{sun2020sparse}. If the sparse arrays are designed such that the peak sidelobe level is low \cite{SUN_SPM_Feature_Article_2020},  
compressed sensing algorithms can be utilized to estimate DOA\cite{roos2019compressed,correas2018experimental}. 
Alternatively, the missing elements in the sparse arrays can be first interpolated using techniques like matrix completion\cite{sun20214d,sun2020sparse,zhang2020doa,zhang2021enhanced}, followed by standard DOA estimation algorithms like MUSIC and ESPRIT.
Matrix completion approach exploits the low-rank property of the Hankel matrix formulated by array received signals, and completes the missing elements using iterative algorithms\cite{zhang2020doa,zhang2021enhanced}. However, typical algorithms for low-rank Hankel matrix completion such as singular value thresholding (SVT) \cite{cai2010singular} has high computational cost due to the compact singular value decomposition (SVD) in each iteration. In \cite{cai2019fast}, an iterative hard thresholding (IHT) algorithm and its accelerated counterpart fast iterative hard thresholding (FIHT) algorithm were proposed. Both IHT and FIHT feature a simple implementation that utilizes efficient methods for SVD computation and Hankel matrix multiplication during the calculations, and FIHT can converge linearly in specific conditions. However, IHT and FIHT require an appropriate initialization and careful parameter tuning to achieve satisfactory estimates.

Benefiting from the rise of deep learning, deep neural networks (DNNs) with various architectures have been proposed for low-rank matrix completion and show superior performance compared with traditional algorithms\cite{fan2017deep,monti2017geometric}. However, these DNNs are usually composed of too many neural layers which leads to a large number of parameters. Furthermore, these DNNs are purely data-driven so they need a huge amount of training data to achieve desirable estimates, which is not available in the scenarios where data collection is expensive.

In this paper, we propose a novel deep learning-based data completion method for sparse array interpolation termed as IHT-Net, and then apply it for DOA estimation. IHT-Net is constructed following the iteration process in IHT algorithm, but set parameters as learnable. In addition, autoencoders structures are introduced to substitute truncated-SVD (t-SVD) operation in IHT algorithm. The autoencoders with multiple linear layers can catch low rank presentations of the signal in the training  process, which serves the same purpose as t-SVD. With extensive numerical simulations, we empirically show that the trained IHT-Net outperforms model-based methods such as FIHT algorithm for both signal reconstruction and DOA estimation using single snapshot.

\section{System model}

A sparse linear array's antenna positions can be considered a subset of a uniform linear array (ULA) antenna positions. Without loss of generality, let the antenna positions of an $M$-element ULA be $\left \{ kd \right \}$, $k=0,1,\cdots,M-1$, where $d=\frac{\lambda }{2}$ is the element spacing with wavelength $\lambda$. Assume there are $P$ uncorrelated far-field target sources in the same range Doppler bin. The impinging signals on the ULA antennas are corrupted by additive white Gaussian Noise with variance of $\sigma^2$. For the single-snapshot case, only the data collected from a single instance in time is available, resulting in the discrete representation of the received signal from a ULA as
\begin{align}
    \mathbf{x} =  \mathbf{A}\mathbf{s} +  \mathbf{n},
    \label{signal_model}
\end{align}
where $\mathbf{x} = \left[x_{1},x_{2},\dots,x_{M}\right]^{T}$, $ \mathbf{A} = \left[\mathbf{a}\left(\theta_{1}\right),\mathbf{a}\left(\theta_{2}\right)\dots,\mathbf{a}\left(\theta_{P}\right)\right]^{T}$, with 
\begin{align}
    \mathbf{a}\left(\theta_{k}\right) = \left [1,e^{j2\pi \frac{d\sin\left(\theta_{k}\right)}{\lambda}},\dots , e^{j2\pi \frac{\left (M-1\right)d\sin \left(\theta_{k}\right)}{\lambda}}\right]^{T},
\end{align}
for $k = 1,\cdots,P$ and $\mathbf{n} = \left[n_{1},n_{2},\dots,n_{M}\right]^{T}$. Then, a Hankel matrix denoted as ${\mathcal{H}}\left( {{{\bf{x}}}} \right) \in {\mathbb C}^{n_1 \times n_2}$
where ${n_{1} + n_{2}} = {M} + 1$,  can be constructed from $\mathbf{x}$ \cite{heinig2011fast}. The Hankel matrix ${\mathcal{H}}\left( {\bf{x}} \right)$ admits a Vandermonde decomposition  structure \cite{Vandermonde_factorization_Hankel_2018,sun20214d,cai2019fast}, i.e.,
\begin{align}
    {\mathcal{H}}\left({{{\bf{x}}}}\right) = {{\bf{V}}_1}{\bf{\Sigma V}}_2^T, 
\end{align}
where ${{\bf{V}}_{1}} = \left[ {{\bf{v}}_{1}\left( {{\theta_1}} \right), \cdots ,{\bf{v}}_{1}\left( {{\theta _P}} \right)} \right]$, ${{\bf{V}}_{2}} = \left[ {{\bf{v}}_{2}\left( {{\theta _1}} \right), \cdots ,{\bf{v}}_{2}\left( {{\theta _P}} \right)} \right]$ with
\begin{align}
{\bf{v}}_{1}\left( {{\theta _k}} \right) &= {\left[ {1,{e^{j2\pi \frac{{d\sin \left( {{\theta _k}} \right)}}{\lambda }}},\cdots, {e^{j2\pi \frac{{\left( {{n_1} - 1} \right)d\sin \left( {{\theta _k}} \right)}}{\lambda }}}} \right]^T}, \\
{\bf{v}}_{2}\left( {{\theta_k}} \right) &= {\left[ {1,{e^{j2\pi \frac{{d\sin \left( {{\theta _k}} \right)}}{\lambda }}},\cdots, {e^{j2\pi \frac{{\left( {n_2 - 1} \right)d\sin \left( {{\theta _k}} \right)}}{\lambda }}}} \right]^T},
\end{align}
and ${\bf{\Sigma }} = {\rm{diag}}\left( {\left[ {{\sigma_1}, {\sigma_2},\cdots ,{\sigma _P}} \right]} \right)$. Assuming that $P\le\min\left ( n_{1},n_{2}\right)$, and both ${\bf{V}}_{1}$ and ${\bf{V}}_{2}$ are full rank of matrices, the rank of the Hankel matrix ${\mathcal{H}}\left( {\bf{x}} \right)$ is indeed $P$, thereby indicating that ${\mathcal{H}}\left( {\bf{x}} \right)$ has low-rank property\cite{cai2019fast}. It's worth noting that a good choice for Hankel matrix size is $n_1  \approx n_2$\cite{chen2013spectral}. This ensures that the resulting matrix ${\mathcal{H}}\left( {\bf{x}} \right)$ is either a square matrix or an approximate square matrix. Specifically, in this paper, we adopt $n_1 = n_2 = \left( {\frac{{M + 1}}{2}} \right) $ if $M$ is odd, and $n_1 = n_2 - 1 = \left( {\frac{{M}}{2}} \right) $ if $M$ is even.

We utilize a 1D virtual SLA synthesized by MIMO radar techniques \cite{SUN_SPM_Feature_Article_2020} with $M_{t}$ transmit antennas and $M_{r}$ receive antennas. The SLA has $M_{t}M_{r} < M$ elements while retaining the same aperture as ULA. Denote the array element indices of ULA as the complete set $\left \{ 1,2,\cdots,M \right \}$, the array element indices of SLA can be expressed as a subset ${\Omega}\subset\left \{ 1,2,\cdots ,M \right \}$. Thus, the signals received by the SLA can be viewed as partial observations of $\mathbf{x}$, and can be expressed as ${\bf{x}}_{s} = {\bf{m}}_{\Omega}\odot\mathbf{x}$, where ${{\bf{m}}_{\Omega}}= {\left[ {{m_1},{m_2}, \cdots ,{m_M}} \right]^T}$ is a masking vector with $m_{j} = 1$, if $j\in\Omega$ or $m_{j} = 0$ if $j\notin\Omega$, and $\odot$ denotes Hadamard product.

\begin{figure*}[htb]
\centering
\includegraphics[width=5.5 in]{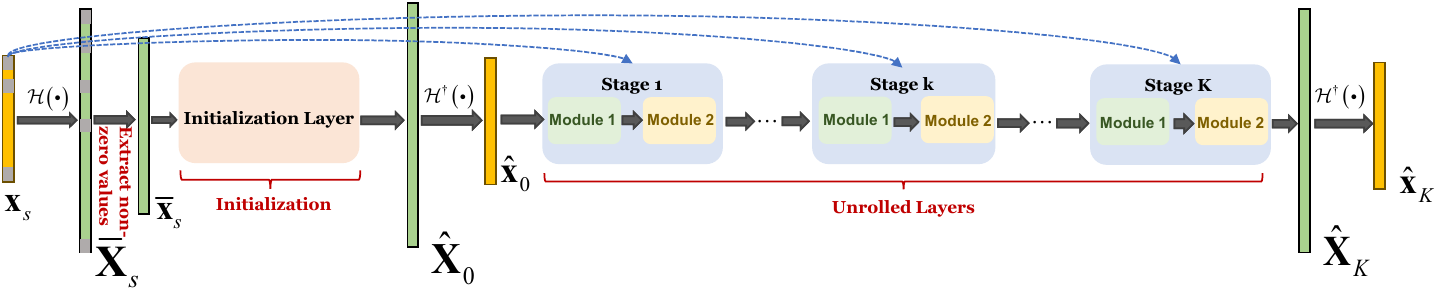}
  \caption{Illustration of IHT-Net architecture}
  \label{IHT-Net}
\end{figure*}
Given the aforementioned statements, the Hankel matrix associated with an SLA configuration can be viewed as a subsampled version of $\mathcal{H}\left({{{\bf{x}}}}\right)$, wherein anti-diagonal entries corresponds to the elements of ${\bf{x}}_{s}$ has values, while the remaining entries are zeros. With the low-rank structure we mentioned before, the missing elements can be recovered by finding the minimum rank of a Hankel matrix that aligns with the known entries\cite{vandereycken2013low}. 
 \begin{align}
  \min_{\bf{x}} \; \mathrm{rank}{\left(\mathcal{H}\left({{\bf{x}}}\right)\right)} \quad
  {\text{s}}{\text{.t}}{\text{. }} \; {\left [ {\mathcal{H}\left({{\bf{x}}}\right)} \right ]_{ij}} = {\left [ {\mathcal{H}\left({{{\bf{x}}_{S}}}\right)} \right ]_{ij}}, \;\left( {i,j} \right) \in \Theta .\label{exact_mc}
\end{align}
Here, $\Theta$ is the set of indices of observed entries that is determined by the SLA. Noting that the rank minimization optimization in (\ref{exact_mc}) is generally an NP-hard problem\cite{vandereycken2013low}. In \cite{cai2019fast} Cai et al. developed iterative hard thresholding (IHT) algorithm for low rank Hankel matrix completion. The convergence speed of IHT algorithm is accelerated by incorporating tangent space projection, resulting in a more expedited variant termed fast iterative hard thresholding (FIHT) algorithm. The main steps in the $i$-th iteration of IHT algorithm are as follows
\begin{align}
  {\bf{X}}_{i} &= \mathcal{H}\left({\bf{x}}_{i} + \beta\left({\bf{x}}_{s} - {\bf{x}}_{i}\right )\right),\label{gradient update}\\
  {\mathbf{x}}_{i+1} &= \mathcal{H}^{\dagger} \left(\mathcal{T}_{r}\left ({\bf{X}}_{i}\right ) \right),\label{hard-thresholding}
\end{align}
where (\ref{gradient update}) is gradient descent update for current estimate ${\mathbf{x}}_{i}$ with fixed step size $\beta$. Then $\mathcal{H}\left(\cdot\right)$ operator transform signal vector from $M\times1$ Euclidean space to an $n_1\times n_2$ Riemannian manifold. Thus, step ($\ref{gradient update}$) can be regarded as one-step gradient descent on a Riemannian manifold\cite{vandereycken2013low}. In step ($\ref{hard-thresholding}$), $\mathcal{T}_{r}$ represents t-SVD for ${\bf{X}}_{i}$, which projects ${\bf{X}}_{i}$ onto the fixed-rank manifold to derive a low rank approximation of ${\bf{X}}_{i}$. Specifically, it is defined as 
\begin{align}
 \mathcal{T}_{r}\left({\bf{X}}_{i}\right )  = \sum_{k=1}^{r}\sigma_{k}{\bf{u}}_{k}{\bf{v}}_{k}^{\ast }, \; \;\sigma_{1}\ge \sigma_{2}\ge\dots \ge\sigma_{r},
\end{align}
where $r$ is the rank of matrix ${\bf X}_i$. The operator $\mathcal{H}^{\dagger}\left ( \cdot  \right ) $ in (\ref{hard-thresholding}) is the inverse of $\mathcal{H}\left ( \cdot  \right ) $ which maps an $n_1 \times n_2$ Hankel matrix to an $M \times 1$ vector. The IHT algorithm runs in an iterative way and has fast convergence speed\cite{cai2019fast}. 
However, achieving optimal results in IHT requires careful parameter tuning (e.g., step size $\beta$ and rank $r$) and can be computationally expensive due to the t-SVD $\mathcal{T}_{r}$ calculations, particularly in scenarios with large Hankel matrix dimensions.

Once the full array response is obtained, DOA can be estimated using high-resolution DOA estimation algorithms that work for single-snapshot \cite{Candes_Dantzig_2007,Liao_Single_Snapshot_MUSIC}.

\section{IHT-Net for low rank Hankel matrix completion}
In order to take advantage of the merits of IHT algprithm and network-based methods, IHT-Net maps the IHT update steps to a deep network architecture that consists of a fixed number of phases, each mirroring one traditional IHT iteration. As shown in Fig. \ref{IHT-Net}, the IHT-Net mainly contains two components: initialization layer, and unrolled layers. The first component provides an initial estimate, analogous to IHT's initialization step, while the second component comprises multiple unrolled layers, mirroring the core iterative steps of the IHT algorithm.
\begin{figure*}
\centering
\subfigure[]{\includegraphics[height=0.9 in, width=3.45 in]{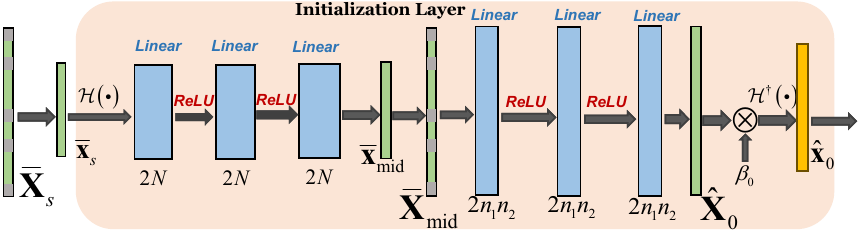}}
\subfigure[]{\includegraphics[height=0.9 in, width=3.51 in]{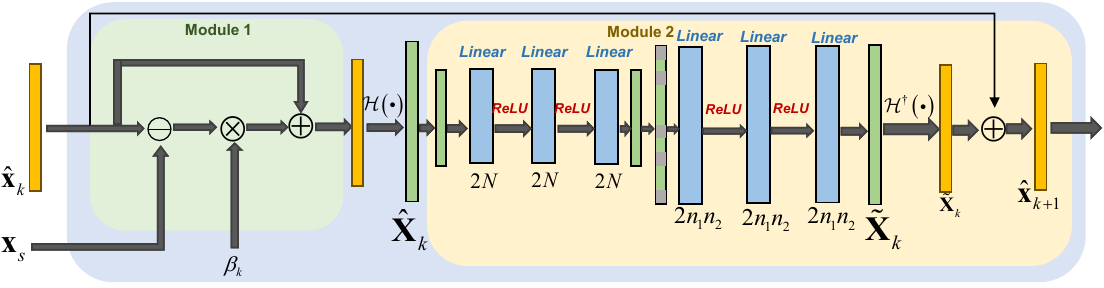}}
 \caption{ (a) Illustration of initialization layer of IHT-Net; (b) Illustration of the $k$th unrolled layer of IHT-Net.}
\label{Layer modules}
\end{figure*}

\subsection{Initialization Layer}
We replace t-SVD in the IHT algorithm with shallow layers autoencoder structures, avoiding the need for the matrix rank knowledge and SVD computation. Inspired by the amazing performance of the masked autoencoders \cite{he2022masked}, we adopt the idea to implement an asymmetric structure that allows an encoder to operate only on observed values (without mask tokens) in the input Hankel vector and a decoder that reconstructs the full signal from the latent representation with mask tokens. As Fig. \ref{IHT-Net} shows, the gray squares represent mask tokens, with the values set to zeros. Since ${\bf{x}}_{s}$ is in complex domain, we concatenate its real part and imaginary part along one dimension and get a $2M \times 1$ vector. This makes the corresponding Hankel vector twice the original length, resulting in a size of $2n_1n_2 \times 1$. Next, non-zeros values are extracted from the Hankel vector $\bar{{\bf{X}}}_{s}$ with dimension $2N \times 1$, denoted as $\bar{{\bf{x}}}_{s}$, along with their positions in the vector denote as a list ${\phi}$, shared across all layers. As mentioned in \cite{jing2020implicit}, inserting multiple extra linear layers in deep neural networks works as implicitly rank minimization of the latent coding. Motivated by this concept, the designed encoder combines 3 linear layers (with bias) separated by rectified linear units (ReLUs). As illustrated in Fig.\ref{Layer modules}(a), the three linear layers share identical input and output dimension $2N$, aligning with the input vector's length. In our implementation, $2N$ is decided by the number of elements in $\Theta$ expressed in (\ref{exact_mc}). We denote the encoder in this layer as $\mathcal{F}_{1}^{\left ( 0\right ) }\left ( \cdot \right)$, so the output of the encoder is defined as  
\begin{align}
 \bar{{\bf{x}}}_{mid}  = \mathcal{F}_{1}^{\left ( 0\right ) }\left(\bar{{\bf{x}}}_{s}\right).
\end{align}

Then, the output $\bar{{\bf{x}}}_{mid}$ is embedded into a $2n_1n_2 \times 1$ zero vector according to the non-zero values positions in original Hankel vector $\bar{{\bf{X}}}_{s}$. The vector is denoted as $\bar{{\bf{X}}}_{mid}$ with dimension $2n_1n_2 \times 1$. For the decoder, we follow the same design pattern as the encoder but with input and output size set to $2n_1n_2$. Denoting the decoder in this layer as $\mathcal{F}_{2}^{\left ( 0\right ) }\left ( \cdot \right)$, then the final output of the initialization layer is 
\begin{align}
 \hat{{\bf{X}}}_{0}  = \beta_0 \mathcal{F}_{2}^{\left ( 0\right ) }\left ( \mathcal{F}_{1}^{\left ( 0\right ) }\left(\bar{{\bf{x}}}_{s}\right) \right)
\end{align}
where $\beta_0$ is a learnable scalar, and parameters of $\mathcal{F}_{1}^{\left ( 0\right ) }\left ( \cdot \right)$, $\mathcal{F}_{2}^{\left ( 0\right ) }\left ( \cdot \right)$ are all learnable. Finally, the Hankel inverse mapping $\mathcal{H}^{\dagger}\left(\cdot\right)$ operates on the output $\hat{{\bf{X}}}_{0}$ to obtain a $2M \times 1$ signal vector $\hat{{\bf{x}}}_{0}$, which is expressed as $\hat{{\bf{x}}}_{0}  = \mathcal{H}^{\dagger}\left(\hat{{\bf{X}}}_{0}\right)$.

\subsection{Unrolled Layers}
The $k$-th unrolled stage consists of two modules. Module 1 is referred to as the Gradient Descent Module, while Module 2 is termed the Low-Rank Approximation Module, which are shown in Fig. \ref{Layer modules}.

The Gradient Descent Module corresponds to Eq.(\ref{gradient update}) in IHT algorithm. With the input $\hat{{\bf{x}}}_{k}$ from the $\left ( k-1 \right ) $-th stage, and ${\bf{x}}_s$ which is broadcasted to every unrolled stage, the intermediate recovery result in $k$-th stage can be defined as
\begin{align}
 {\hat{\bf{X}}}_{k}  = \mathcal{H}\left({\hat{{\bf{x}}}_{k} + \beta_k\left ( {\bf{x}}_{s} - {\hat{\bf{x}}}_{k} \right )} \right),
\end{align}
where the step size $\beta_k$ is a learnable parameter.

The Low-Rank Approximation Module keeps totally the same architecture as the initialization layer, while introducing a skip connection within the layers. We firstly extract ${\hat{\bf{x}}}_{k}$ from the output ${\hat{\bf{X}}}_{k}$ according to the non-zero values position list ${\phi}$. Then the output of this module is derived by passing it through the autoencoders, resulting in 
\begin{align}
 \tilde{{\bf{X}}}_{k}  = \mathcal{F}_{1}^{\left ( k\right ) }\left ( \mathcal{F}_{2}^{\left ( k\right ) }\left(\hat{{\bf{x}}}_{k}\right) \right).
\end{align}
After Hankel inverse operation, we have $ \tilde{{\bf{x}}}_{k}  =\mathcal{H}^{\dagger}\left(\tilde{{\bf{X}}}_{k}\right)$.
With the skip connection between the input ${\hat{\bf{x}}}_{k}$ and the output $\tilde{{\bf{x}}}_{k}$, the final output of the $k$-th unrolled stage is
\begin{align}
 \hat{{\bf{x}}}_{k+1}  = \tilde{{\bf{x}}}_{k} + \gamma_{k}\left (\tilde{{\bf{x}}}_{k} - \hat{{\bf{x}}}_{k} \right ) 
\end{align}
where $\gamma_{k}$ is a learnable parameter weighting the residual term $\left (\tilde{{\bf{x}}}_{k} - \hat{{\bf{x}}}_{k} \right)$. The final estimate $\hat{{\bf{x}}}_{K}$ is obtained after $K$ unrolled stages forward inference.

\subsection{IHT-Net Training Specifics}
\label{sec_training}

We generate $P$ point-target sources in the same range-Doppler bin. The angles of the sources follow a uniform distribution within the field of view (FoV) spanning $\left [ -60^{\circ} ,60^{\circ} \right ]$. Their amplitudes have a uniform distribution ranging from $\left [0.5,1\right ]$, while their phases are uniformly distributed between $\left [0,2\pi \right ]$. Following equation (\ref{signal_model}), we can generate $N_{b}$ training labels without noise, denoted as $\left \{ {\bf{x}}_{label}^{q} \right \} _{q=1}^{N_{b}}$ for specific SLA configuration. Then we obtain inputs $\left \{ {\bf{x}}_{input}^{q}\right \} _{q=1}^{N_{b}}$ for network training by adding
different levels of Gaussian white noise with SNR randomly chosen from $\left [ 10\mathrm{dB} ,30\mathrm{dB}\right ]$ for each training sample. In our experiment, $P=2$, $N_{b} = 700000$.

The learnable parameters in $k$-th phase of IHT-Net are $\left \{\beta_{k},\gamma_{k},\mathcal{F}_{1}^{\left ( k\right)},\mathcal{F}_{2}^{\left(k\right)}\right\}$. Hence, the
learnable parameter set of IHT-Net is $\left \{\beta_{k},\gamma_{k},\mathcal{F}_{1}^{\left ( k\right)},\mathcal{F}_{2}^{\left(k\right)}\right\}_{k=1}^{K}$. When $k=0$, the learnable parameters are $\left \{\beta_{0},\mathcal{F}_{1}^{\left(0\right)},\mathcal{F}_{2}^{\left(0\right)}\right\}$.

Given the training data pairs $\left \{ {\bf{x}}_{label}^{q},{\bf{x}}_{input}^{q}\right \}_{q=1}^{N_{b}}$ (Note ${\bf{x}}_{label}^{q}$ and ${\bf{x}}_{input}^{q}$ are block data which has $N$ training samples pairs), IHT-Net 
first takes ${\bf{x}}_{input}^{q}$ as input and generates the output as the reconstruction results denoted as $\hat{{\bf{x}}}_{K}^{q}$. We aim to reduce the discrepancy between $\hat{{\bf{x}}}_{K}^{q}$ and ${\bf{x}}_{label}^{q}$, while satisfying the low-rank approximation constraint, which can be stated as $\mathcal{F}_{1} \circ \mathcal{F}_{2} \approx \mathcal{I}$. Therefore, the loss function for IHT-Net is designed as follows
\begin{align}
 {Loss}_{total}  = {Loss}_{1} + \alpha{Loss}_{2}
\end{align}
with
\begin{align}
 {Loss}_{1}  &= \frac{1}{N_{b}N}\sum_{q=1}^{N_{b}} \left \|\hat{{\bf{x}}}_{K}^{q} - {\bf{x}}_{label}^{q}\right \|_{2}^{2}, 
 \label{loss1}
\\
 {Loss}_{2}  &= \frac{1}{N_{b}N}\sum_{q=1}^{N_{b}}\sum_{k=0}^{K} \left \|\mathcal{H}^{\dagger}\left(\mathcal{F}_{1}^{\left ( k\right ) }\left ( \mathcal{F}_{2}^{\left ( k\right ) }\left(\hat{{\bf{x}}}_{k}^{q}\right) \right)\right) - \hat{{\bf{x}}}_{k}^{q}\right \|_{2}^{2}, 
\end{align}
where $K+1$, $\alpha$ are the total number of IHT-Net phases and the regularization parameter, respectively. In our experiments, $\alpha$ is set to $0.01$.
For training IHT-Net,  Adam optimization algorithm\cite{kingma2014adam} is employed with an initial learning rate of $10^{-4}$, which decays to 0.5 times of the original rate every 10 epochs.
\begin{figure*}
\centering
\subfigure[]{\includegraphics[width=1.80 in]{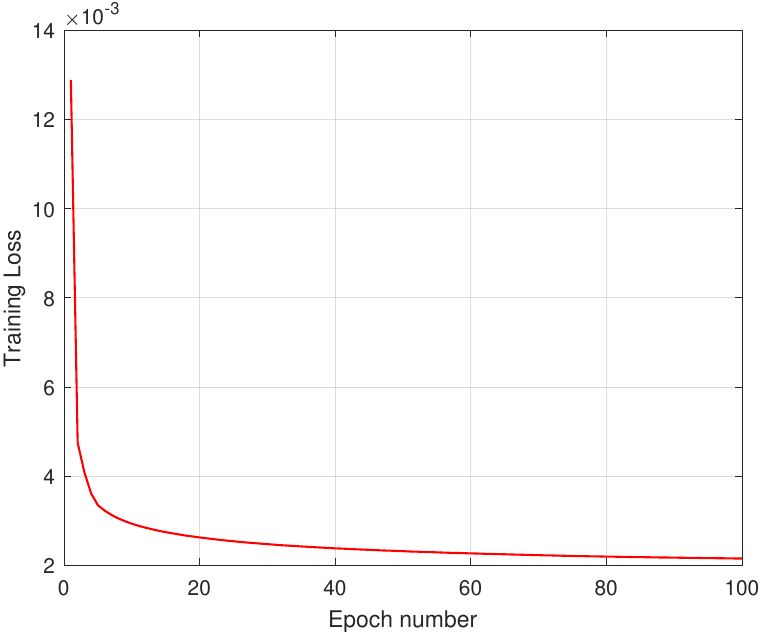}}
\subfigure[]{\includegraphics[width=1.75 in]{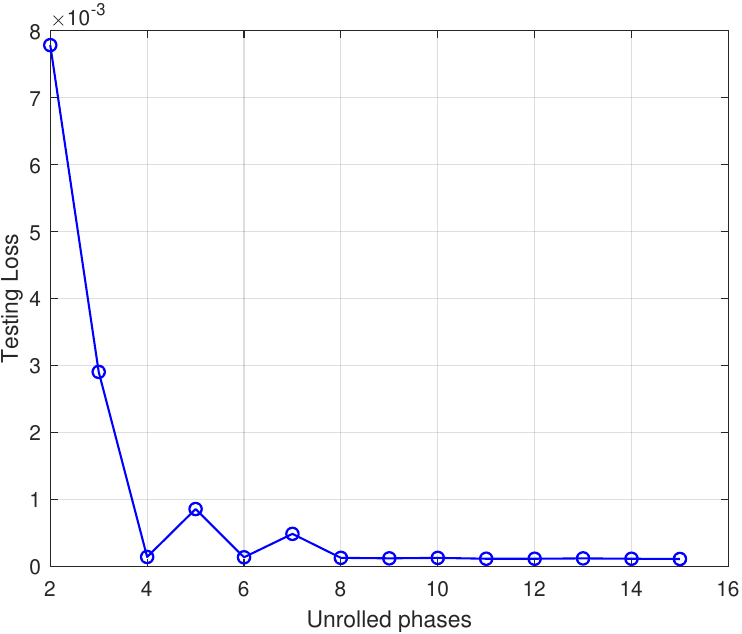}}
\subfigure[]{\includegraphics[width=1.80 in]{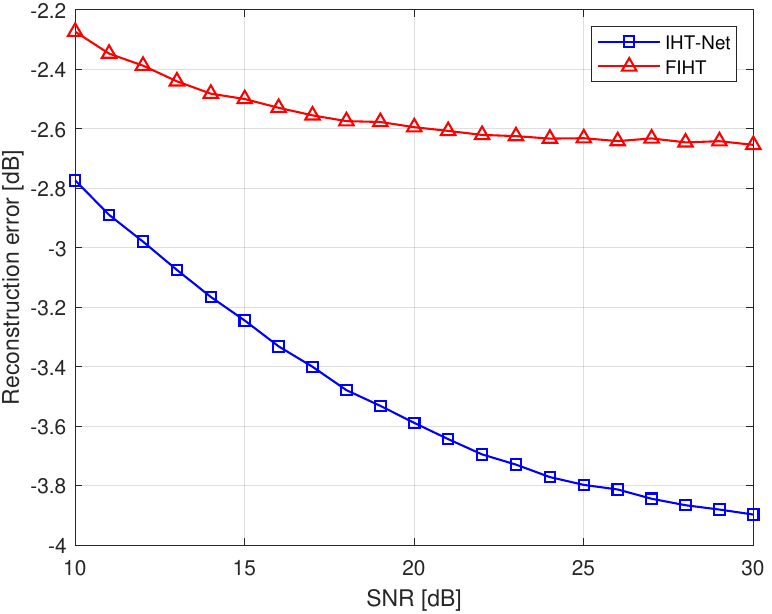}}
 \caption{ (a) IHT-Net training loss (defined in (\ref{loss1})) v.s. epoch for IHT-Net with 8 unrolled phases; (b) IHT-Net testing loss (defined in (\ref{loss1})) with various numbers of unrolled phases of IHT-Net; (c) Signal reconstruction error comparison between IHT-Net and FIHT\cite{cai2019fast} in different SNRs.}
\label{testing loss comparison}
\end{figure*}
\begin{figure*}
\centering
\subfigure[]{\includegraphics[width=1.70 in]{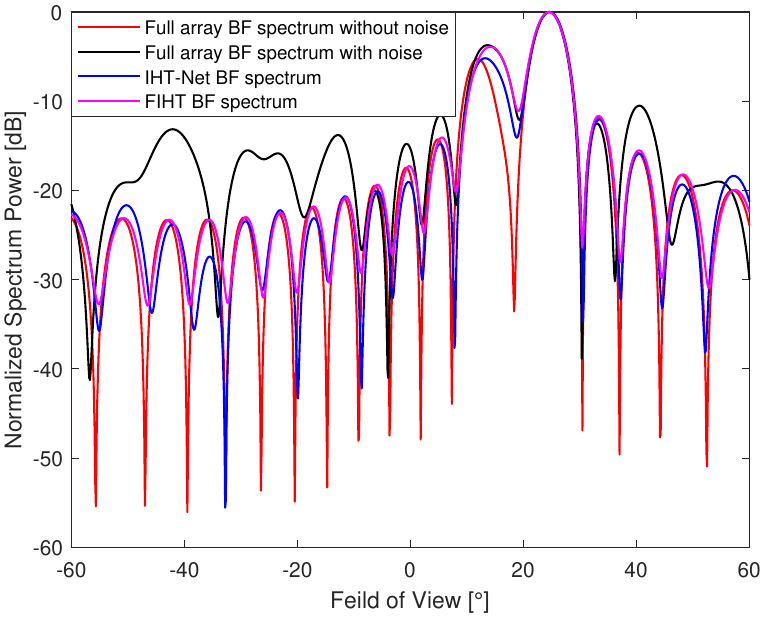}}
\subfigure[]{\includegraphics[width=1.70 in]{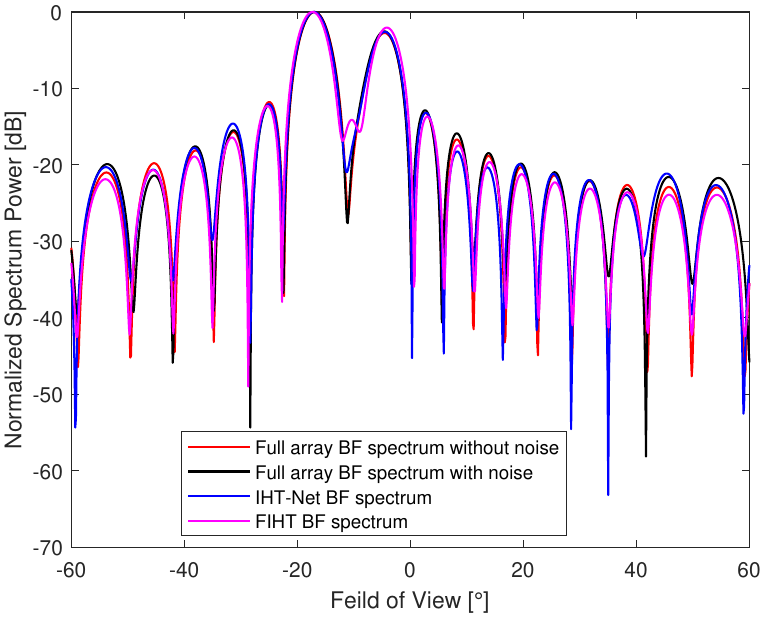}}
\subfigure[]{\includegraphics[width=1.70 in]{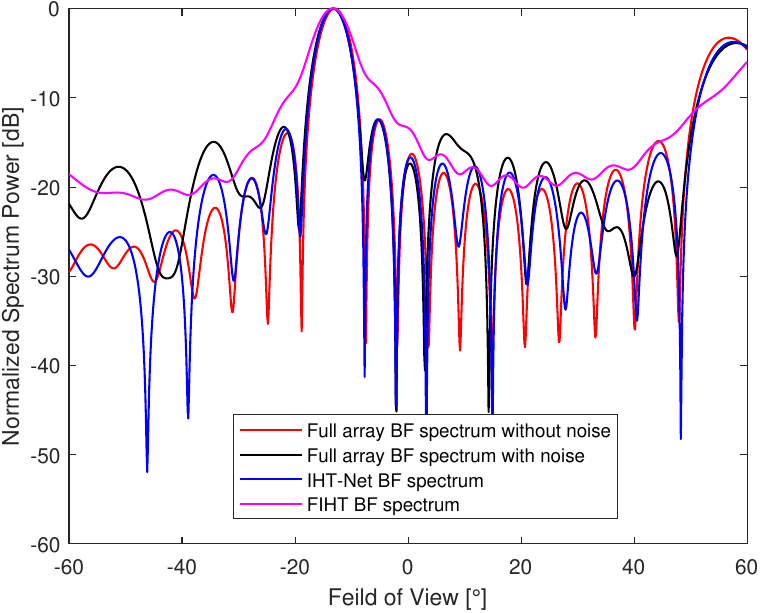}}
\subfigure[]{\includegraphics[width=1.70 in]{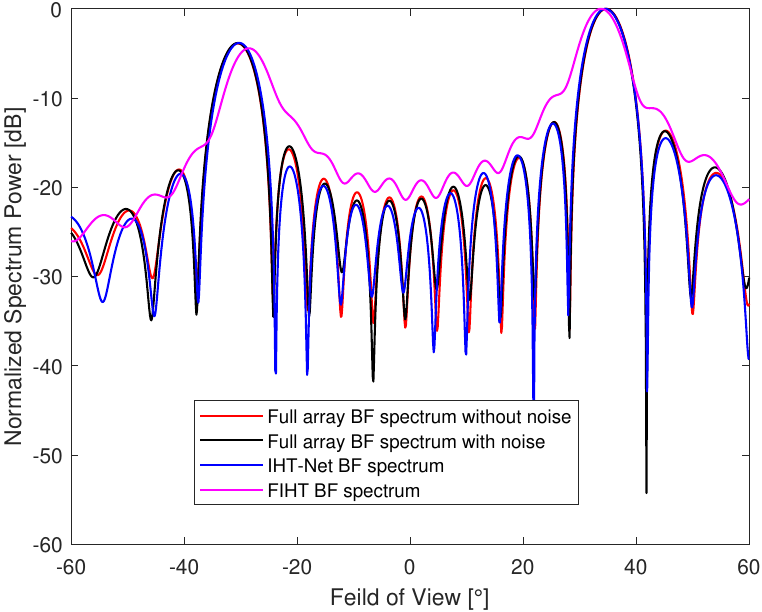}}
 \caption{Beamforming spectrum examples in different SNRs with different SLAs; (a) SNR=10dB, 18-element SLA; (b) SNR=30dB, 18-element SLA; (c) SNR=10dB,  10-element SLA; (d) SNR=30dB, 10-element SLA.}
\label{testing spectrum comparison}
\end{figure*}
\section{Numerical Results}
In this section, we evaluate IHT-Net's performance via numerical simulations. A ULA with $N=21$ elements is considered, and an SLA is derived from the 21-element ULA by randomly choosing part of its antennas. We first perform experiments using an 18-element SLA, with the training dataset generated following the strategy described in Section \ref{sec_training}. Total $100$ epochs of training are conducted in our training procedure. Fig. \ref{testing loss comparison} (a) shows the initial rapid decay of the training loss within the first 10 epochs, which indicates that the proposed IHT-Net is easily trainable. To verify the recovery performance of IHT-Net with different layers, we trained IHT-Net with different layers using the same training datasets. Then we randomly generated $5,000$ testing samples in 20dB SNR for evaluation. The testing loss is calculated as (\ref{loss1}). Fig. \ref{testing loss comparison} (b) shows that the testing loss decreases as the number of unrolled phases increases, but this decline stabilizes after 8 phases. Thus, we choose to utilize 8 unrolled phases to balance reconstruction performance and computational efficiency in IHT-Net.

Furthermore, we compared IHT-Net and the FIHT algorithm\cite{cai2019fast} in different SNRs, using a testing dataset of $5,000$ samples per SNR level. In Fig. \ref{testing loss comparison} (c), IHT-Net consistently outperforms FIHT in reconstruction loss, particularly at higher SNR, highlighting its superior performance. We also conducted experiments employing a 10-element SLA, and compared the recovered spectrums in various SNRs with different SLAs. Fig. \ref{testing spectrum comparison} explicitly shows the beam patterns of recovered full array response by IHT-Net and FIHT, as compared with the spectrums of full array response with and without noise. 
It can be found that the proposed IHT-Net has denoising ability in relatively low SNR, e.g. 10dB, which indicates that the modules in IHT-Net play the same role as the t-SVD operation in FIHT. In addition, both IHT-Net and FIHT obtain promising spectrums in high SNR e.g. 30dB. Fig. \ref{testing spectrum comparison}(c) and (d) illustrate that for a sparser SLA, FIHT struggles to recover the original signal effectively. In contrast, IHT-Net consistently produces satisfactorily recovered spectrums which keep the mainlobes and sidelobes, confirming its superior recovery performance, particularly with sparser SLAs.
\begin{figure}
\centering
\includegraphics[width=2.10in]{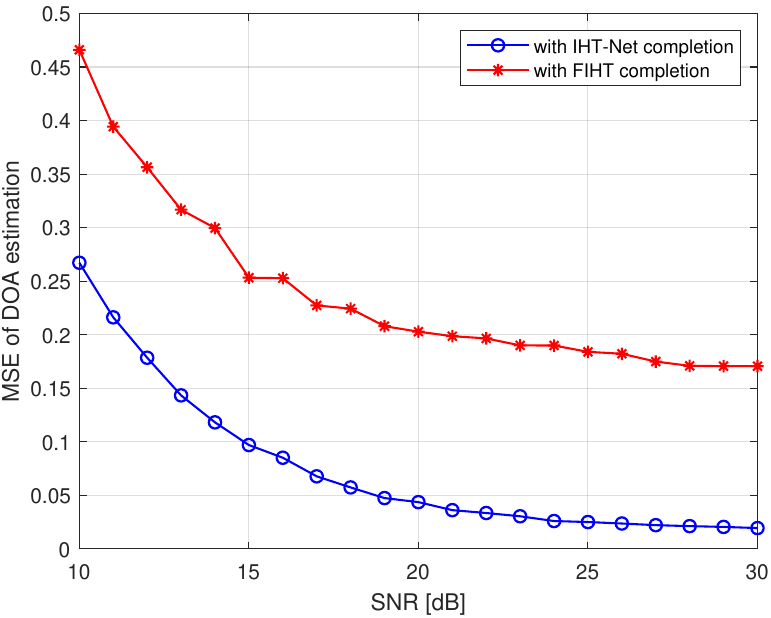}
\caption{Comparison of DOA estimation errors after IHT-Net and FIHT\cite{cai2019fast} completion under different SNRs.}
\label{testing_loss}
\vspace{-2mm}
\end{figure}

Finally, we compared the mean square errors of DOA estimation using IHT-Net and FIHT reconstruction in different SNRs. We employed beamforming (BF) for DOA estimation. The testing samples number in each SNR is also $5,000$. The errors were calculated using mean-square-loss (MSE). The results shown in Fig. \ref{testing_loss} demonstrated that the IHT-Net completion leads to improved DOA estimation accuracy compared with FIHT completion.

\section{Conclusions}
We have demonstrated a novel learning-based sparse array interpolation approach for single-snapshot DOA estimation, termed IHT-Net. It holds potential for applications in automotive radar systems employing sparse arrays. Derived from the FIHT algorithm, IHT-Net incorporates learnable parameters and nonlinear layers, offering an enhanced optimizer through supervised learning with shallow unrolled layers. IHT-Net is easily trainable and interpretable, facilitating further network design and development. Numerical simulations demonstrate its superior reconstruction and DOA estimation performance compared with FIHT.

 \newpage
\bibliographystyle{IEEEtran}

{\small
\balance
\bibliography{refs}
}

\end{sloppypar}
\end{document}